PHYSYCS OF STARS
AND INTERSTELLAR MEDIUM

# Determining the Effective Temperatures of G- and K-Type Giants and Supergiants Based on Observed Photometric Indices

L. S. Lyubimkov and D. B. Poklad

*Crimean Astrophysical Observatory, Nauchny, Crimea, 298409 Russia*
*e-mail: lyub@crao.crimea.ua*
Received October 7, 2013

**Abstract**—A method for determining the effective temperature $T_{eff}$ of G- and K-type giants and supergiants is proposed. The method is based on the use of two photometric indices free from the interstellar absorption influence: the $Q$ index in the $UBV$ photometric system and the $[c_1]$ index in the $uvby$ system. Empirical relations between the $T_{eff}$ values found for nearby and bright G- and K-type giants and supergiants with the use of the infrared fluxes method (IRFM), on the one hand, and the observed $Q$ and $[c_1]$ indices for these stars, on the other hand, are plotted. A systematic discrepancy between the dependences of $T_{eff}$ on $Q$ for the stars with standard and reduced metallicities is found. Approximating the plotted dependences with second-order polynomials, one can obtain a relatively simple and a rather accurate method for determining the $T_{eff}$ value in the range of $3800 \leq T \leq 5100$ K (based on the $Q$ index) or $4900 \leq T_{eff} \leq 5500$ K (based on the $[c_1]$ index).

**DOI:** 10.3103/S0884591314050055

## INTRODUCTION

The chemical composition of F-, G-, and K-type giants and supergiants has attracted particular attention of researchers for decades. This may be attributed to the fact that anomalous abundances of a number of light elements are observed in the atmospheres of such stars. These anomalies are associated with the changes in the chemical composition of a star in the process of its evolution [2]. Specifically, nitrogen tends to be overabundant, and carbon tends to be deficient. The $^{12}C/^{13}C$ carbon isotope ratio turns out to be reduced greatly. The lithium abundance varies greatly [12]. It follows from the calculations that deep convective mixing takes place when stars reach the evolutionary stages of F-, G-, and K-type giants and supergiants. This mixing results in the expulsion of the products of thermonuclear reactions from the stellar interior towards the surface. An accurate analysis of the observed chemical anomalies and their comparison with the theoretical predictions allow one to verify and refine the modern stellar evolution theory.

A number of studies of A-, F-, and G-type supergiants and bright giants have already been conducted at the Crimean Astrophysical Observatory [12–14]. We plan to continue these investigations and study the chemical composition of a large group of cooler G- and K-type giants and supergiants (luminosity classes I, II, III). It is commonly known that the analysis of the chemical composition of a star starts with the determination of a pair of fundamental parameters: the effective temperature $T_{eff}$ and the free-fall acceleration $\log g$ in the atmosphere. The accuracy of the determined element abundances is highly dependent on the accuracy of these parameters. The first one (effective temperature $T_{eff}$) is of particular importance to our task.

When a large number of stars are studied, it is important that a relatively simple and a rather accurate method suitable for en masse determination of the effective temperature $T_{eff}$ should be available. A direct method for the estimation of $T_{eff}$ by measuring the angular diameters of stars (e.g., in their occultation by the Moon) is known. The authors of [20] used this method to obtain high-precision $T_{eff}$ values for 32 K- and M-type giants. However, this method is rather complex and is not suitable for the estimation of $T_{eff}$ values of stars from an arbitrarily given sample.

When other methods for determining $T_{eff}$ are discussed, it should be noted that these methods exhibit certain peculiarities when applied to cool G- and K-type stars. Specifically, the $H_\beta$ and $H_\gamma$ Balmer lines that serve as reliable indicators of $T_{eff}$ in the case of hotter A- and F-type stars are not suitable for G- and K-type stars due to the weakness and strong blending of these lines in the spectra.

It is known that the infrared flux (IRF) method proposed more than 30 years ago [6] is one of the most accurate methods for determining $T_{eff}$ of cool stars. This method is based on the fact that the radiation fluxes of cool stars are rather large in the infrared region of the spectrum; besides, the infrared region contains a relatively small number of lines and is free from any significant interstellar absorption. The IRF method has been applied already to a large number of cool stars.

In a departure from the purely photometrical IRF method, the authors of [1, 11] proposed a spectroscopic method for determining $T_{eff}$ of F-, G-, and K-type giants and supergiants that uses the ratios of depths of pairs of lines with widely different excitation potentials as $T_{eff}$ indicators. The dependences of the mentioned ratios on $T_{eff}$ are calibrated based on the known high-precision determinations of $T_{eff}$. This method was used to determine the $T_{eff}$ values for 110 F-, G-, and K-type giants [1] and 161 supergiants of the same spectral types [11]. Since this method requires high-precision measurements of depths of a number of spectral lines (or, more specifically, pairs of lines), the spectra with a sufficiently high resolution should be available in order for the method to be applicable.

We have determined $T_{eff}$ and $\log g$ and a number of other parameters for 63 galactic A-, F-, and G-type supergiants and bright giants (luminosity classes I, II) in [14]. Stellar parallaxes, whose observed values were taken from the new reduction of the HIPPARCOS satellite data [22], were used to determine $\log g$ in this study. Such a method for determining $\log g$ is described in [3, 14] and allows one to obtain the desired values with unprecedented precision; for example, the mean error in $\log g$ values for the stars with distances of up to 300 pc equaled ±0.06 dex. As for $T_{eff}$, a complex technique that combined spectroscopy (Balmer lines) and photometry (the $Q$, $[c_1]$, and $\beta$ indices) was applied in the case of A- and F-type stars. We were forced to use only the photometric indices in the case of cooler G-type stars; besides, the $\beta$ index was excluded due to the fact that its use resulted in a significant overestimation of $T_{eff}$ with respect to the $[c_1]$ index and the IRF method data.

The $[c_1]$ index and the parallax from [22] served as the primary indicators of $T_{eff}$ and $\log g$, respectively, for G-type supergiants and giants in [14]. The observed $[c_1]$ index was compared directly to the theoretical $[c_1]$ values calculated based on the atmospheric models. When cooler K-type stars ($T_{eff}$ < 4900 K) are considered, the theory, as it will be shown below, does not provide a satisfactory explanation of the observed $[c_1]$ and $Q$ values. The present study relies on an approach different to the one used in [14]: we plotted the empirical dependences of high-precision $T_{eff}$ values determined using the IRF method on the observed $Q$ and $[c_1]$ indices for nearby and bright stars. Approximating the plotted dependences with second-order polynomials, we obtained a relatively simple and a rather accurate method for estimating the $T_{eff}$ parameter for G- and K-type giants and supergiants.

## $Q$ AND $[C_1]$ INDICES: COMPARISON OF OBSERVATIONS WITH THEORY

The photometric indices considered in the present study are defined in the following way: $Q = (U - B) - 0.72(B - V)$ in the $UBV$ photometric system and $[c_1] = c_1 - 0.20(b - y)$ in the $uvby$ system. It is known that both these indices are free from the influence of interstellar absorption.

It is interesting to see how the theoretical $Q$ and $[c_1]$ indices compare to their observed values for cool G- and K-type stars. We list below the $T_{eff}$ values for many G- and K-type giants and supergiants obtained using the IRF method below; note that we found their $\log g$ values to be varying from 1.2 to 3.0. In order to conduct a correct comparison with the theory, we selected only the stars with $\log g = 1.6-2.4$ (or $\log g = 2.0 \pm 0.4$) from this data array and compared them to the calculations for $\log g = 2.0$. In addition to this, only the stars with a standard (solar) metallicity (i.e., [Fe/H] ≈ 0.0) were included into the comparison. The results of the comparison between observations and theory are shown in Fig. 1.

Figure 1a shows the observed $Q$ values for stars with their effective temperatures $T_{eff}$ ranging from 4000 to 5500 K and $\log g = 2.0 \pm 0.4$ (dots) in comparison with the $Q$ values calculated using the ATLAS software [7] (solid line). The observed dependence of $Q$ on $T_{eff}$ is approximated with a dashed curve; this curve reveals a discrepancy between the observations and calculations at $T_{eff}$ < 4500 K that gets larger as $T_{eff}$ is reduced. If the theoretical dependence is used directly in the determination of $T_{eff}$ based on the observed $Q$ index for, e.g., a K-type giant with $T_{eff}$ = 4200 K, the effective temperature $T_{eff}$ is reduced by 200 K. If one takes into account the sensitivity of the spectra of such cool stars to the $T_{eff}$ parameter, the mentioned discrepancy becomes significant. The systematic underestimation of $T_{eff}$ for K-type giants and supergiants may result in systematic errors in the determination of element abundances.

It can be seen from Fig. 1b that more significant discrepancies are observed in the case of the $[c_1]$ index. Two theoretical curves (solid lines) were obtained based on the data provided by two research groups that used different software (ATLAS [7] and MARCS [17]). The theoretical curves deviate from each other markedly at $T_{eff}$ < 5500 K. The curve obtained with the use of the ATLAS software agrees well with the observed dependence of $T_{\text{eff}}$ on $[c_1]$ (dashed curve) down up to $T_{\text{eff}} = 4500$ K; however, a large discrepancy between the observations and theory is seen at $T_{\text{eff}} < 4500$ K. It is interesting that the theoretical curve obtained using the MARCS software lies notably above both the observed dependence and the theoretical ATLAS curve at $T_{\text{eff}} < 5500$ K. The region of $T_{\text{eff}}$ temperatures ranging from 4000 to 4950 K and $[c_1]$ values ranging from 0.2 to 0.3 is of importance for further analysis. This region is shaded in Fig. 1b. The observed $[c_1]$ index reveals no substantial dependence on $T_{\text{eff}}$ in the shaded area; therefore, $[c_1]$ is not suitable for the determination of $T_{\text{eff}}$ in this region.

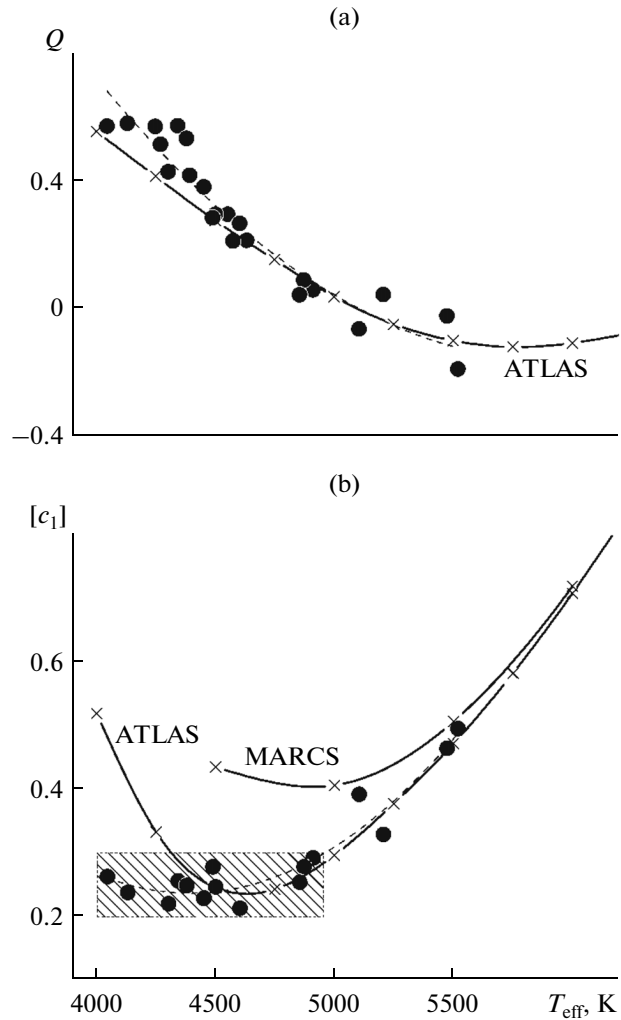

**Fig. 1.** Dependences of the observed $Q$ and $[c_1]$ indices for stars with $\log g = 2.0 \pm 0.4$ (dots) on the effective temperature $T_{\text{eff}}$. Solid curves correspond to the results of calculations done using the ATLAS [7] and MARCS [17] software for $\log g = 2.0$. Dashed curves represent the averaged dependences plotted based on the observed values. The area where the observed $[c_1]$ index reveals no substantial dependence on $T_{\text{eff}}$ is shaded.

Thus, firstly, the calculated $Q$ and $[c_1]$ indices do not provide a satisfactory description of the observed $Q$ and $[c_1]$ values for K-type giants and supergiants at temperatures $T_{\text{eff}} < 4500$ K. Therefore, it is incorrect to compare the observed $Q$ and $[c_1]$ values to the calculated indices with the aim of determining the effective temperature at $T_{\text{eff}} < 4500$ K. Another approach is proposed in the present study: we apply the observed $Q$ and $[c_1]$ indices directly by inserting them into the obtained dependences of $T_{\text{eff}}$ on $Q$ and $[c_1]$. Secondly, an analysis of Fig. 1 allows us to specify the $T_{\text{eff}}$ intervals where the obtained dependences may produce fairly reliable estimates of $T_{\text{eff}}$. We are of the opinion that the dependence of $T_{\text{eff}}$ on $Q$ may be used for relatively cool stars at $T_{\text{eff}} < 5100$ K where the variation of $Q$ with $T_{\text{eff}}$ is rather steep (Fig. 1a), while the dependence of $T_{\text{eff}}$ on $[c_1]$ may, as follows from what was said above, be used only for hotter stars with $T_{\text{eff}} > 4900$ K (Fig. 1b). These dependences overlap in the region of $T_{\text{eff}} = 4900-5100$ K and may provide the means for determining $T_{\text{eff}}$ in the entire temperature region occupied by G- and K-type giants and supergiants.

This brings up the question of discrepancies between the theory and observations at $T_{eff} < 4500$ K. It follows from the calculations presented in [7] that convection assumes a significant role at such low temperatures even in the case of giants and supergiants (a similar effect in cool dwarfs is very well known). Therefore, the method used to take this convection into account in the calculations of model atmospheres also becomes significant. It may be assumed that the traditional mixing length theory used in such calculations becomes untenable at $T_{eff} < 4500$ K. It is necessary to use three-dimensional nonstationary atmospheric models (see, for example, [21]).

## SELECTION OF STARS WITH HIGH-PRECISION $T_{EFF}$ VALUES OBTAINED USING THE INFRARED FLUXES METHOD

The photometric indices are, in contrast to the spectroscopic indicators of the effective temperature $T_{eff}$, relatively easily obtained from observations and are well suited for estimating $T_{eff}$ even in the case of distant stars when a high-resolution stellar spectrum may not be measured. We plot the empirical dependences of $T_{eff}$ on $Q$ and $[c_1]$ based on high-precision $T_{eff}$ values obtained for nearby G- and K-type giants and supergiants with the use of the infrared fluxes method (IRFM). The data (effective temperatures $T_{eff}$ determined using the IRFM) from [4, 5, 19] are used in the present study. These three papers contain the data on a fairly large number of G- and K-type giants and supergiants. The $T_{eff}$ values listed in these papers were obtained with high precision; besides, it follows from the comparison of $T_{eff}$ for shared stars that no significant systematic discrepancies between these studies are present.

The selection process described below resulted in the retrieval of 81 stars from [4, 5, 19] for further analysis. These objects were divided into two groups (stars with normal and reduced metallicities). This is explained by the fact that giants and supergiants with reduced metallicities differed systematically in their observed $Q$ indices from normal stars with the same $T_{eff}$ temperatures. Table 1 lists the data for 65 selected stars with normal metallicities and 16 stars with reduced metallicities. Normal metallicities are defined as the values of [Fe/H] ranging from −0.2 to +0.2 ([Fe/H] = 0.0 on average), while the [Fe/H] values of objects with reduced metallicities vary from −0.3 to −0.7 ([Fe/H] = −0.5 on average). The stars with their metallicities falling out of the range from −0.7 to +0.2 were not analyzed. Since all the considered stars are fairly bright, several estimates of [Fe/H] are usually given for these stars in scientific literature. These estimates may be found, for example, in the SIMBAD database (http://simbad.u-strasbg.fr/simbad/sim-fid). It bears reminding that the [Fe/H] value characterizes the iron abundance with respect to the solar iron abundance on a logarithmic scale.

It can be seen from Table 1 that the selected stars are bright: their apparent stellar magnitudes $m_V < 5.6$. The condition that they should have fairly large and high-precision parallaxes π was used as another selection criterion. The fulfillment of this condition is required in order to determine the log$g$ value with certainty (see below). The parallax selection resulted in the inclusion of nearby stars with distances $d < 200$ pc into our list. An exception is found in the case of the late F-type giant HR 9057. This star is the hottest in our list ($T_{eff} = 5520$ K) and has $m_V < 6.03$ and $d = 323$ pc.

The stars selected from the above-mentioned papers [4, 5, 19] had such $T_{eff}$ values that cover the entire region of $T_{eff}$ occupied by G- and K-type giants and supergiants. Table 1 lists the stars with $T_{eff}$ ranging from 3850 to 5520 K. You will recall that these values were determined using the IRFM and are, thus, fairly accurate. The errors in $T_{eff}$ determination given in [4, 5, 19] allow us to conclude that the typical error in the values of $T_{eff}$ listed in Table 1 is ±(60−80) K.

The free-fall acceleration log$g$ values that should be appropriate for giants and supergiants were also taken into account in the process of selection of stars. All the selected stars fall within the log$g$ range of 1.2–3.0. The log$g$ values were determined based on the stellar parallaxes π expressed in milliarcseconds (mas). The method for determining log$g$ based on parallaxes is described in [3, 14]. It was already noted above that this method provides an unprecedented accuracy of the log$g$ determination (several hundredths) for nearby stars similar to the ones selected in our study. The determination of log$g$ also requires estimating the mass $M$ of each star; this problem was solved with the use of evolutionary tracks [8] for stars with normal metallicities and tracks [9] for stars with reduced metallicities. It should be noted that the use of tracks [9] instead of tracks [8] results in a certain reduction in log$g$ and a notable reduction in the value of $M$. Table 1 shows that the masses $M$ of the stars under consideration vary from $0.8 M_\odot$ to $7.6 M_\odot$, where $M_\odot$ is the solar mass.

Another essential condition applied in the selection of stars from [4, 5, 19] consisted in the availability of the observed $Q$ index. The $Q$ values were obtained using the catalogue [16] and the SIMBAD database (http://simbad.u-strasbg.fr/simbad/sim-fid). It turned out that the observed $[c_1]$ index values (taken from [10]) are not known for all the selected stars. The observed $Q$ and $[c_1]$ values are listed in the two rightmost columns of Table 1.

**Table 1.** Parameters of the selected stars with normal and reduced metallicities ([Fe/H] ≈ 0.0 and −0.5)

| HR | HD | Sp | $m_V$ | $\pi$, mas | $d$, pc | $T_{\rm eff}$, K (IRFM) | Source | $\log g$ | $M/M_\odot$ | $Q$ | $[c_1]$ |
|---|---|---|---|---|---|---|---|---|---|---|---|
| | | | | | | [Fe/H] ≈ 0.0 | | | | | |
| 165 | 3627 | K3 III | 3.28 | 30.91 | 32 | 4343 | [19] | 2.22 | 1.46 | 0.568 | 0.256 |
| 168 | 3712 | K0 III | 2.25 | 14.29 | 70 | 4553 | [19] | 1.85 | 5.51 | 0.290 | |
| 180 | 3919 | G8 III | 4.61 | 13.27 | 75 | 5100 | [5] | 2.78 | 3.10 | 0.026 | |
| 253 | 5234 | K2 III | 4.83 | 9.93 | 101 | 4453 | [4] | 2.17 | 2.57 | 0.376 | 0.229 |
| 402 | 8512 | K0 III | 3.60 | 28.66 | 35 | 4689 | [4] | 2.67 | 2.17 | 0.164 | 0.287 |
| 434 | 9138 | K4 III | 4.84 | 10.73 | 93 | 4046 | [4] | 1.66 | 1.41 | 0.566 | 0.263 |
| 464 | 9927 | K3 III | 3.57 | 18.41 | 54 | 4380 | [19] | 2.10 | 2.26 | 0.528 | 0.249 |
| 489 | 10380 | K3 III | 4.45 | 8.98 | 111 | 4132 | [19] | 1.64 | 2.27 | 0.575 | 0.238 |
| 603 | 12533 | K3 II | 2.14 | 9.19 | 109 | 4259 | [19] | 1.35 | 7.56 | 0.594 | |
| 617 | 12929 | K2 III | 2.00 | 49.56 | 20 | 4501 | [19] | 2.37 | 2.03 | 0.289 | 0.247 |
| 694 | 14770 | G8 III | 5.19 | 8.17 | 122 | 4951 | [5] | 2.57 | 3.54 | 0.052 | |
| 874 | 18322 | K1 III | 3.89 | 23.89 | 42 | 4608 | [4] | 2.55 | 2.05 | 0.200 | 0.273 |
| 941 | 19476 | K0 III | 3.80 | 28.93 | 35 | 4879 | [4] | 2.90 | 2.42 | 0.124 | 0.285 |
| 1318 | 26846 | K3 III | 4.90 | 13.46 | 74 | 4577 | [4] | 2.47 | 2.15 | 0.301 | 0.267 |
| 1457 | 29139 | K5 III | 0.98 | 48.94 | 20 | 3883 | [19] | 1.32 | 1.55 | 0.793 | 0.182 |
| 2012 | 39003 | G9.5 III | 3.95 | 14.16 | 71 | 4604 | [4] | 2.31 | 3.20 | 0.261 | 0.213 |
| 2427 | 47174 | K3 Iab | 4.80 | 7.84 | 128 | 4394 | [5] | 1.99 | 3.07 | 0.412 | |
| 2443 | 47442 | K0 II–III | 4.43 | 7.74 | 129 | 4633 | [5] | 2.13 | 4.34 | 0.208 | |
| 2985 | 62345 | G8 III | 3.57 | 23.07 | 43 | 5001 | [4] | 2.79 | 3.02 | 0.019 | 0.281 |
| 2990 | 62509 | K0 III | 1.15 | 96.54 | 10 | 4833 | [19] | 2.88 | 2.28 | 0.135 | 0.298 |
| 3003 | 62721 | K4 III | 4.88 | 9.61 | 104 | 3988 | [19] | 1.54 | 1.46 | 0.711 | 0.281 |
| 3475 | 74739 | G8 Iab | 4.03 | 9.85 | 102 | 4911 | [4] | 2.32 | 4.25 | 0.052 | 0.292 |
| 3547 | 76294 | G9 II–III | 3.13 | 19.51 | 51 | 4817 | [19] | 2.45 | 3.66 | 0.075 | |
| 3705 | 80493 | K7 III | 3.16 | 16.06 | 62 | 3851 | [19] | 1.23 | 1.77 | 0.827 | |
| 3903 | 85444 | G7 III | 4.12 | 12.36 | 81 | 5085 | [5] | 2.57 | 3.59 | −0.022 | |
| 3994 | 88284 | K0 III | 3.61 | 28.98 | 35 | 4865 | [4] | 2.84 | 2.54 | 0.188 | |
| 4247 | 94264 | K0 III | 3.83 | 34.38 | 29 | 4670 | [19] | 2.80 | 1.69 | 0.171 | 0.279 |
| 4291 | 95345 | K1 III | 4.85 | 9.05 | 110 | 4490 | [4] | 2.18 | 2.91 | 0.278 | 0.278 |
| 4392 | 98839 | G7.5 III | 4.99 | 6.12 | 163 | 4872 | [4] | 2.29 | 4.29 | 0.083 | 0.278 |
| 4432 | 99998 | K3.5 III | 4.77 | 5.40 | 185 | 3919 | [19] | 1.19 | 2.67 | 0.701 | 0.171 |
| 4716 | 107950 | G6 III | 4.77 | 8.44 | 118 | 5033 | [4] | 2.50 | 3.76 | −0.017 | |
| 4932 | 113226 | G8 III | 2.83 | 29.76 | 34 | 5049 | [19] | 2.74 | 3.17 | 0.054 | |
| 5429 | 127665 | K3 III | 3.58 | 20.37 | 49 | 4271 | [4] | 2.00 | 1.77 | 0.509 | |
| 5480 | 129312 | G7 III | 4.86 | 6.07 | 165 | 4854 | [4] | 2.23 | 4.46 | 0.037 | 0.254 |
| 5649 | 134505 | G7 III | 3.41 | 27.80 | 36 | 5058 | [5] | 2.88 | 2.84 | −0.001 | 0.291 |
| 6132 | 148387 | G8 III | 2.74 | 35.42 | 28 | 5007 | [5] | 2.82 | 2.94 | 0.024 | |
| 6147 | 148786 | G9 III | 4.29 | 13.39 | 75 | 5106 | [5] | 2.69 | 3.32 | 0.049 | |
| 6603 | 161096 | K2 III | 2.75 | 39.85 | 25 | 4533 | [4] | 2.47 | 1.89 | 0.397 | 0.305 |
| 6698 | 163917 | G9 III | 3.31 | 21.64 | 46 | 4871 | [4] | 2.60 | 3.30 | 0.156 | |
| 6703 | 163993 | G8 III | 3.70 | 23.84 | 42 | 5011 | [5] | 2.84 | 2.88 | 0.011 | |
| 6705 | 164058 | K5 III | 2.23 | 21.14 | 47 | 3927 | [19] | 1.28 | 2.20 | 0.782 | 0.185 |
| 6770 | 165760 | G8 III | 4.65 | 11.96 | 84 | 4969 | [4] | 2.66 | 3.33 | 0.047 | 0.311 |
| 6807 | 166640 | G8 III | 5.57 | 8.28 | 121 | 5079 | [5] | 2.75 | 3.17 | −0.001 | |

**Table 1.** (Contd.)

| HR | HD | Sp | $m_V$ | $\pi$, mas | $d$, pc | $T_{eff}$, K (IRFM) | Source | $\log g$ | $M/M_\odot$ | $Q$ | $[c_1]$ |
|---|---|---|---|---|---|---|---|---|---|---|---|
| 6895 | 169414 | K2 III | 3.84 | 27.42 | 36 | 4450 | [19] | 2.43 | 1.50 | 0.322 | 0.226 |
| 6970 | 171391 | G8 III | 5.13 | 9.91 | 101 | 5116 | [5] | 2.75 | 3.19 | −0.066 | |
| 6973 | 171443 | K3 III | 3.85 | 16.38 | 61 | 4248 | [4] | 1.93 | 1.89 | 0.565 | |
| 7259 | 178345 | K0 II | 4.12 | 6.88 | 145 | 4575 | [5] | 1.94 | 5.07 | 0.206 | |
| 7328 | 181276 | G9 III | 3.80 | 26.27 | 38 | 4935 | [19] | 2.89 | 2.62 | 0.048 | 0.313 |
| 7429 | 184406 | K3 III | 4.45 | 30.31 | 33 | 4473 | [4] | 2.63 | 1.10 | 0.399 | 0.275 |
| 7479 | 185758 | G1 II | 4.39 | 7.67 | 130 | 5415 | [5] | 2.46 | 3.91 | −0.136 | 0.370 |
| 7581 | 188114 | K0 II–III | 4.13 | 17.94 | 56 | 4683 | [5] | 2.55 | 2.61 | 0.127 | 0.236 |
| 7615 | 188947 | K0 III | 3.88 | 24.17 | 41 | 4796 | [4] | 2.76 | 2.47 | 0.145 | 0.281 |
| 7754 | 192947 | G8.5 III–IV | 3.58 | 30.82 | 32 | 4978 | [4] | 2.96 | 2.54 | 0.009 | 0.296 |
| 7949 | 197989 | K0 III | 2.48 | 44.86 | 22 | 4710 | [19] | 2.67 | 2.32 | 0.116 | 0.300 |
| 8093 | 201381 | G8 III | 4.52 | 20.47 | 49 | 5093 | [5] | 3.03 | 2.55 | 0.018 | 0.274 |
| 8167 | 203387 | G8 III | 4.30 | 16.58 | 60 | 5105 | [5] | 2.84 | 2.98 | −0.063 | |
| 8232 | 204867 | G0 Ib | 2.91 | 6.07 | 165 | 5474 | [5] | 1.87 | 5.99 | −0.029 | 0.464 |
| 8255 | 205512 | K1 III | 4.88 | 14.09 | 71 | 4609 | [4] | 2.52 | 2.17 | 0.229 | 0.241 |
| 8414 | 209750 | G2 Ib | 2.95 | 6.23 | 161 | 5206 | [5] | 1.81 | 6.26 | 0.038 | 0.329 |
| 8498 | 211388 | K3 II–III | 4.13 | 5.25 | 190 | 4140 | [5] | 1.39 | 4.99 | 0.610 | |
| 8632 | 214868 | K2 III | 4.51 | 9.80 | 102 | 4303 | [19] | 1.92 | 2.53 | 0.423 | 0.220 |
| 8649 | 215167 | K3 III | 4.69 | 7.53 | 133 | 4072 | [4] | 1.51 | 2.38 | 0.567 | 0.185 |
| 8650 | 215182 | G2 II–III | 2.95 | 15.22 | 66 | 5104 | [5] | 2.36 | 4.16 | −0.070 | 0.392 |
| 8916 | 220954 | K1 III | 4.28 | 21.96 | 46 | 4699 | [4] | 2.70 | 2.13 | 0.223 | 0.273 |
| 9057 | 224342 | F8 III | 6.03 | 3.10 | 323 | 5520 | [4] | 2.35 | 4.32 | −0.196 | 0.495 |
| | | | | | [Fe/H] $\approx$ −0.5 | | | | | | |
| 163 | 3546 | G8 III | 4.37 | 19.91 | 50 | 4935 | [19] | 2.75 | 1.91 | −0.161 | 0.342 |
| 1726 | 34334 | K2.5 III | 4.55 | 14.04 | 71 | 4193 | [4] | 1.74 | 0.81 | 0.357 | 0.220 |
| 1907 | 37160 | K0 III | 4.09 | 27.76 | 36 | 4693 | [4] | 2.57 | 1.16 | −0.062 | 0.301 |
| 2035 | 39364 | K1 III–IV | 3.81 | 28.68 | 35 | 4599 | [4] | 2.41 | 1.11 | −0.031 | 0.318 |
| 3403 | 73108 | K2 III | 4.60 | 12.74 | 78 | 4387 | [4] | 1.94 | 1.25 | 0.303 | — |
| 4382 | 98430 | K0 III | 3.56 | 17.56 | 57 | 4468 | [4] | 1.98 | 1.66 | 0.172 | — |
| 4518 | 102224 | K0.5 III | 3.71 | 17.76 | 56 | 4378 | [19] | 1.90 | 1.34 | 0.302 | 0.254 |
| 4608 | 104979 | G8 III | 4.13 | 19.98 | 50 | 4824 | [4] | 2.57 | 1.84 | −0.075 | 0.187 |
| 5340 | 124897 | K1.5 III | −0.04 | 88.83 | 11 | 4231 | [19] | 1.66 | 1.21 | 0.382 | 0.337 |
| 5681 | 135722 | G8 III | 3.47 | 26.78 | 37 | 4834 | [19] | 2.58 | 1.86 | −0.012 | 0.289 |
| 5787 | 138905 | K0 III | 3.92 | 19.99 | 50 | 4711 | [4] | 2.40 | 1.73 | 0.017 | 0.275 |
| 5889 | 141714 | G3.5 III | 4.60 | 19.18 | 52 | 5247 | [5] | 3.02 | 2.25 | −0.211 | 0.313 |
| 6220 | 150997 | G7.5 III | 3.49 | 30.02 | 33 | 4948 | [19] | 2.77 | 1.93 | −0.068 | — |
| 6869 | 168723 | K0 III–IV | 3.26 | 53.93 | 19 | 4835 | [4] | 2.88 | 1.18 | −0.025 | 0.307 |
| 8551 | 212943 | K0 III | 4.79 | 21.99 | 45 | 4588 | [4] | 2.51 | 0.90 | 0.125 | 0.296 |
| 8961 | 222107 | G8 III | 3.82 | 37.87 | 26 | 4605 | [19] | 2.60 | 0.84 | −0.034 | 0.283 |

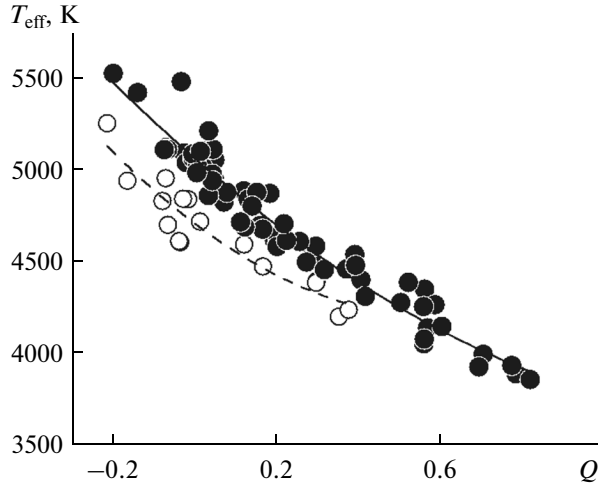

**Fig. 2.** Relation between the effective temperature $T_{\text{eff}}$ and the $Q$ index for stars with normal metallicities (filled circles) and reduced metallicities (open circles). Solid and dashed curves correspond to relations (1) and (2), respectively.

## RELATIONS BETWEEN THE EFFECTIVE TEMPERATURE AND THE OBSERVED $Q$ AND $[c_1]$ INDICES

The authors of [15] plotted the calibration dependences of $T_{\text{eff}}$ on $Q$ and $[c_1]$ based on the data obtained for 102 early- and medium-type main-sequence B stars. These dependences allowed one to determine promptly the effective temperature $T_{\text{eff}}$ for stars of the indicated type with known observed values of $Q$ and $[c_1]$. Virtually the same concept is applied in the present study: we plot the dependences of $T_{\text{eff}}$ on the $Q$ and $[c_1]$ indices. However, in contrast to the authors of [15], we do not use our own $T_{\text{eff}}$ estimates and rely on high-precision $T_{\text{eff}}$ values obtained with the use of the IRF method in other studies for nearby and bright stars.

Figure 2 shows the relation between $T_{\text{eff}}$ and $Q$ plotted based on the data from Table 1. Filled and open circles correspond to stars with normal ($[\text{Fe/H}] \approx 0.0$) and reduced ($[\text{Fe/H}] \approx -0.5$) metallicities, respectively. We approximated both dependences with second-order polynomials (see solid and dashed curves in Fig. 2, respectively). It can be seen that the second dependence (for stars with reduced metallicities) lies systematically below the first one. It was already noted in the discussion of Fig. 1a that the empirical dependence of $T_{\text{eff}}$ on $Q$ should be used for relatively cool stars. We may now define the limits of its applicability more precisely: the obtained $T_{\text{eff}}(Q)$ curves in Fig. 2 may be used to determine the effective temperature in the $T_{\text{eff}}$ interval of 3800–5100 K for stars with normal metallicities (solid curve) and in the $T_{\text{eff}}$ interval of 4200–5100 K for stars with reduced metallicities (dashed curve). The corresponding $Q$ intervals are roughly as follows: 0.9–0.0 in the first case and from 0.4 to –0.2 in the second one.

Thus, the estimation of $T_{\text{eff}}$ based on the $Q$ index may be performed using the following relations:

$$T_{\text{eff}} = 5064.0 - 1942.8Q + 639.59Q^2 \quad ([\text{Fe/H}] \approx 0.0,\ 0.0 \leq Q \leq 0.9), \tag{1}$$

and

$$T_{\text{eff}} = 4708.1 - 1684.9Q + 1353.9Q^2 \quad ([\text{Fe/H}] \approx -0.5,\ -0.2 \leq Q \leq 0.4). \tag{2}$$

As for the $[c_1]$ index, a relatively small number of stars may be used to plot the dependence of $T_{\text{eff}}$ on $[c_1]$. Firstly, the values of $[c_1]$ are, in contrast to the $Q$ values, not known for all the selected stars. Table 1 lists the $[c_1]$ indices for 42 out of 65 objects with normal metallicities and 13 out of 16 objects with reduced metallicities. Secondly, it was already noted above that the $[c_1]$ index becomes insensitive to the effective temperature at $T_{\text{eff}} < 4900$ K. Therefore, we are forced to exclude a significant number of stars with $T_{\text{eff}} < 4900$ K from the analysis. Figure 3 shows ten stars (filled circles) with $T_{\text{eff}}$ ranging from 4854 to 5520 K and normal ($[\text{Fe/H}] \approx 0.0$) metallicities. In addition to these, a compact group of eight giants with their $\log g$ values being maximally close to $\log g = 3$ ($\log g = 2.89 \pm 0.09$ on average) is present. Since these giants have very close $T_{\text{eff}}$ and $[c_1]$ values, we represented them with a mean point at $[c_1] = 0.290 \pm 0.012$ and $T_{\text{eff}} = 4947 \pm 106$ K (open circle) in order to avoid overloading Fig. 3. The approximation of the dependence for ten mentioned stars with a second-order polynomial is shown with a dashed curve, and the solid curve represents the theoretical dependence obtained in calculations with the use of the ATLAS software [7] at $\log g = 2.0$. It can be seen that these curves agree with each other well. In other words, the observed $[c_1]$ index values are well described with a theoretical curve in the $T_{eff}$ temperature interval of 4900–5500 K. It is interesting that this curve corresponding to the free-fall acceleration $\log g = 2.0$ goes through the point with a considerably different value of $\log g = 2.89 \pm 0.09$ (open circle). This suggests that the $[c_1]$ index depends only weakly on $\log g$ at $T_{eff} \approx 5000$ K. However, the calculations show that the dependence may be more pronounced at $T_{eff} \approx 5500$ K.

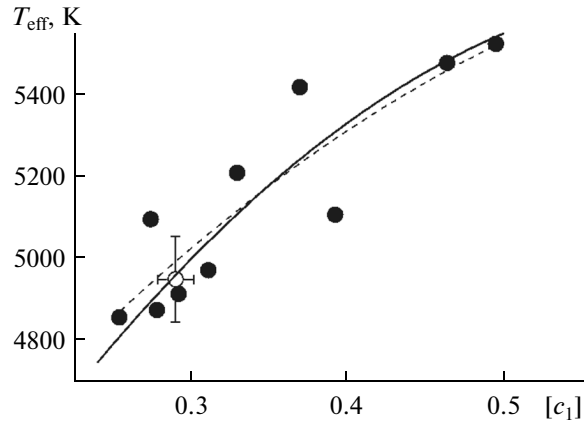

**Fig. 3.** Relation between the effective temperature $T_{eff}$ and the $[c_1]$ index for stars with normal metallicities (filled circles). The open circle at $[c_1] = 0.290 \pm 0.012$ and $T_{eff} = 4947 \pm 106$ K represents eight giants with very close $T_{eff}$ and $[c_1]$ parameters (see text). The dashed curve is an approximation of the observed points with a second-order polynomial. The solid curve represents the results of calculations [7] performed using the ATLAS software at $\log g = 2.0$. This curve corresponds to relation (3).

We arrived at a conclusion that the theoretical dependence represented by the solid curve in Fig. 3 may be used to determine the effective temperature based on the $[c_1]$ index in the interval of $4900 \leq T_{eff} \leq 5500$ K (or $0.27 \leq [c_1] \leq 0.50$). The equation for this curve takes on the following form:

$$T_{eff} = 3325.0 + 7267.5[c_1] - 5648.8[c_1]^2 \qquad (3)$$
$$([Fe/H] \approx 0.0, 0.27 \leq [c_1] \leq 0.50).$$

It should be noted that the mentioned interval of $4900 \leq T_{eff} \leq 5500$ K corresponds to G-type giants and supergiants. Thus, relation (3) is not suitable for cooler K-type stars; when such stars are analyzed, one may use relations (1) and (2). All three relations (1)–(3) may be valid in a common region of $T_{eff}$ values located between 4900 and 5100 K. When analyzing the connection between the $Q$ and $[c_1]$ indices and the effective temperature $T_{eff}$, one should take into account the fact that these indices may also depend to a certain extent on $\log g$. It was already noted that the $\log g$ values for the stars in our list fall within the interval of 1.2–3.0. It is interesting that the value of $\log g = 2$ lies, according to [14], roughly at the border between supergiants (luminosity class I) and giants (luminosity classes II, III). In order to examine the probable connection between the $Q$ and $[c_1]$ indices and $\log g$, we divided all the stars into two groups with $\log g > 2$ and $\log g < 2$. It turned out that the $T_{eff}(Q)$ and $T_{eff}([c_1])$ dependences reveal no systematic differences between these two groups. You will also recall that the point with $\log g = 2.89$ in Fig. 3 practically belonged to the curve calculated at $\log g = 2.0$. Therefore, it may be assumed that relations (1)–(3) are applicable in the entire range of $\log g$ values varying from 1.2 to 3.0.

## ACCURACY OF THE METHOD AND COMPARISON WITH OTHER DATA

How accurate is the determination of $T_{eff}$ with the use of Eqs. (1), (2), and (3)? In order to answer this question, one should first estimate the effect of errors in the $Q$ and $[c_1]$ indices. The errors in the observed values of $Q$ and $[c_1]$ for the stars from Table 1 may be determined based on the scatter of the measured color indices $U - B$, $B - V$, $b - y$, and $c_1$ that are found in the relations defining $Q$ and $[c_1]$ (see above). It can be seen from the SIMBAD database (http://simbad.u-strasbg.fr/simbad/sim-fid) that each of these bright stars generally has several measurements of the mentioned parameters. We found that the $\Delta Q$ error varies from $\pm 0.01$ to $\pm 0.04$ for the majority of stars, and the $\Delta[c_1]$ error varies within the same interval. The errors $\Delta Q = \pm(0.02-0.03)$ and $\Delta[c_1] = \pm(0.02-0.03)$ may be considered typical.

252 LYUBIMKOV, POKLAD

**Table 2.** Comparison of different estimates of the physical parameters of Arcturus

| $T_{eff}$, K | $\log g$ | $M/M_\odot$ | Notes |
| --- | --- | --- | --- |
| 4231 ± 49 | 1.65 ± 0.07 | 1.25 ± 0.32 | $T_{eff}$ — IRFM [19] |
| 4262 ± 20 | 1.69 ± 0.03 | 1.30 ± 0.15 | $T_{eff}$ — relation (2) |
| 4286 ± 30 | 1.66 ± 0.05 | 1.08 ± 0.06 | Data from [18] |

It turned out that the inaccuracies in the determination of $T_{eff}$ from Eq. (1) depend not only on the $\Delta Q$ error but also on the $Q$ value itself. Our analysis showed that the $\Delta T_{eff}$ error is small for the coolest stars with $Q = 0.6$–0.7: $\Delta T_{eff}$ here equals ±(10–30) K at $\Delta Q = \pm(0.02$–0.03). The error for relatively hot stars with $Q \approx -0.1$ is significantly larger: $\Delta T_{eff} = \pm(40$–60) K at $\Delta Q = \pm(0.02$–0.03). It is important that the $\Delta T_{eff}$ value is lower than ±100 K even at $\Delta Q = \pm 0.04$. As for Eq. (2) obtained for stars with reduced metallicities, the $\Delta T_{eff}$ error here turned out to be essentially the same as for Eq. (1) at $Q = -0.1$, but was two times smaller than the error for Eq. (1) at $Q = 0.3$.

When Eq. (3) is applied, the $\Delta T_{eff}$ error is, as in the previous case, dependent not only on the $\Delta[c_1]$ error but also on the $[c_1]$ index itself. It follows from our analysis that the $\Delta T_{eff}$ value is rather small for relatively hot stars with $[c_1] = 0.4$–0.5: $\Delta T_{eff}$ here equals ±(40–80) K at $\Delta[c_1] = \pm(0.02$–0.03). However, the uncertainty increases when the $[c_1]$ index and, correspondingly, the $T_{eff}$ temperature are reduced and reaches ±100 K already at $[c_1] = 0.3$.

It follows from the presented analysis that a fairly high precision in the entire interval of $T_{eff} = 3800$–5100 K recommended above is guaranteed when the effective temperature $T_{eff}$ is determined with the use of the $Q$ index and Eqs. (1) or (2) are applied. The corresponding typical $\Delta T_{eff}$ error is ±50 K or lower. Note that the indicated $T_{eff}$ interval covers the entire region of $T_{eff}$ values occupied by K-type giants and supergiants (roughly from 3800 to 4800 K). In the case of G-type stars (except for the very latest-type ones for which the $Q$ index may still be applied), one may use only the $[c_1]$ index by inserting it into Eq. (3). The latter approach was recommended to be applied in the $T_{eff}$ interval of 4900–5500 K. However, relatively high accuracy is achieved only in the analysis of the hottest stars with $[c_1] \approx 0.4$–0.5 and $T_{eff} \approx 5300$–5500 K. The $\Delta T_{eff}$ error at lower $[c_1]$ values (and lower $T_{eff}$ temperatures) reaches ±100 K and above.

In order to obtain an additional estimate of the accuracy of our method, one may compare the $T_{eff}$ values calculated using Eqs. (1)–(3) to the $T_{eff}$ temperatures found using the IRF method and listed in Table 1. In other words, one may estimate the mean error by looking at the scatter of individual points in Figs. 2 and 3 around the curves defined by Eqs. (1)–(3). We found that the mentioned scatter is characterized by the quadratic deviations of ±94, ±98, and ±115 K in the case of relations (1), (2), and (3), respectively. It should not be overlooked that this scatter is induced in part by the errors in the initial $T_{eff}$ values found using the IRF method. If we assume that the accuracy of the IRF method is ±(60–80) K (see above), the typical error of our method equals roughly ±(50–70) K for Eqs. (1) and (2) and ±(80–100) K for relation (3). These estimates are generally comparable to the estimates obtained above and confirm the fact that the use of the $Q$ index in the determination of $T_{eff}$ for K-type stars actually provides a higher accuracy than the use of the $[c_1]$ index in the determination of $T_{eff}$ for G-type stars.

One may also gain some insight into the accuracy of our method by comparing the obtained estimates with the values of $T_{eff}$ for G- and K-type giants and supergiants determined using the above-mentioned spectroscopic method (based on measuring the ratios of depths of pairs of lines with different excitation potentials) proposed in [1, 11].

Eight stars from our list are also analyzed in [1], and three stars from the list are found in [11]. We determined $T_{eff}$ based on the $Q$ index from Eqs. (1) and (2) for the first eight stars and based on the $[c_1]$ index from Eq. (3) for the remaining three stars. Having compared our $T_{eff}$ values for these 11 shared stars with the data from [1, 11], we found that the difference equals −38 K on average. If a correction for this systematic value is introduced, the differences between the data provided by the two methods are characterized by a quadratic deviation of ±79 K. The errors of both methods contribute to this value. Assuming that the contributions are equal, we obtain an error of ±56 K for each method. It can be seen that even such a simplified analysis for a relatively small number of shared stars verifies the decent accuracy of our method.

Arcturus (HR 5340, a bright and very nearby K-type giant that became the subject of extensive research) was included into the list of stars analyzed in our study. It is commonly known that Arcturus is often viewed as a test object for verifying any methods or results related to the study of cool stars. Following this tradition, we compared our $T_{eff}$, $\log g$, and $M$ values for Arcturus with recent results from [18]. The results of this comparison are presented in Table 2. The data from [18] are given in the last row of the table (it should be noted that the value of [Fe/H] = −0.52 ± 0.04 found in [18] confirms that the star has a reduced metallicity). Note that the $T_{eff}$ parameter was determined in [18] based on the observed flux distribution in a wide spectral range, and the $\log g$ value was found using the exact same method applied in the present study (i.e., with the use of the parallax).

Table 2 lists the values of $T_{eff}$ determined both with the use of the IRF method [19] and our method (specifically, from Eq. (2) based on the observed index $Q = 0.382 \pm 0.018$). The results agree well with the data from [18] in both cases; however, the $T_{eff}$ temperature obtained from Eq. (2) was noticeably closer to the value from [18]. The values of $\log g$ and $M$ agree in both cases with the ones found in [18] to within the limits of estimation error. The primary conclusion that may be drawn from the analysis of Table 2 is as follows: the example of Arcturus verifies the high precision of the determination of $T_{eff}$ with the use of our method.

## CONCLUSIONS

The present study was aimed at developing a relatively simple and a rather accurate method for determining the effective temperature $T_{eff}$ value for cool G- and K-type giants and supergiants. Two photometric indices free from the influence of interstellar absorption were used to achieve this goal: the $Q$ index in the *UBV* system and the $[c_1]$ index in the *uvby* system. Having compared the observed dependence of these indices on the $T_{eff}$ value determined using the IRF method to the theoretical calculations, we found significant discrepancies between the observations and theory at temperatures $T_{eff} < 4500$ K. It was concluded that a direct comparison of the observed $Q$ and $[c_1]$ indices to the theoretical calculations at such low temperatures may result in significant errors in the determination of $T_{eff}$. The results of this comparison become especially unreliable in the case of the $[c_1]$ index at $T_{eff} < 4900$ K. Therefore, we propose to use these indices in another way: the observed $Q$ and $[c_1]$ values should be inserted into the dependences of $T_{eff}$ on $Q$ and $[c_1]$ that were found in our study and are based on the determination of $T_{eff}$ with the use of the IRF method.

The obtained dependences represented by Eqs. (1)–(3) allow one to determine $T_{eff}$ based on the $Q$ index in the region of $3800 \leq T_{eff} \leq 5100$ K and based on the $[c_1]$ index in the region of $4900 \leq T_{eff} \leq 5500$ K. The dependences of $T_{eff}$ on $Q$ for stars with normal ([Fe/H] ≈ 0.0) and reduced ([Fe/H] ≈ −0.5) metallicities turned out to be different; these dependences are represented by relations (1) and (2). Our analysis reveals that the accuracy of the proposed method is comparable to the accuracy of the IRF method. If we assume that the accuracy of the IRF method is ±(60–80) K, the typical error of our method equals roughly ±(50–70) K for Eqs. (1) and (2) and ±(80–100) K for relation (3). The error in determining the $T_{eff}$ value for the coolest bright stars (e.g., Arcturus) may be as low as ±20 K.

*Translated by D. Safin*